\documentclass[11pt]{article}
\usepackage{amsmath, amssymb, amsthm, amscd, amsfonts}
\usepackage[margin=1in]{geometry}
\usepackage{times}
\usepackage[colorlinks=true,linkcolor=blue,citecolor=blue,allcolors=blue]{hyperref}
\hypersetup{
  colorlinks=true,
  linkcolor=blue,
  filecolor=blue,
  urlcolor=blue,
  allcolors=blue
}
\usepackage{import}
\usepackage{cite}

\usepackage{fontspec}
\usepackage{polyglossia}
\setdefaultlanguage{english}
\setotherlanguages{telugu, sanskrit}

\newfontfamily\telugufont{KohinoorTelugu}[
    Path = Fonts/KohinoorTelugu/,
    Extension = .ttf,
    UprightFont = *-Regular,
    BoldFont = *-Bold,
    FontFace = {l}{n}{*-Light},
    FontFace = {m}{n}{*-Medium},
    FontFace = {sb}{n}{*-Semibold},
    AutoFakeSlant=0.2,
]

\newfontfamily\devanagarifont{KohinoorDevanagari}[
    Path = Fonts/KohinoorDevanagari/,
    Extension = .otf,
    UprightFont = *-Regular,
    BoldFont = *-Bold,
    FontFace = {l}{n}{*-Light},
    FontFace = {m}{n}{*-Medium},
    FontFace = {sb}{n}{*-Semibold},
    AutoFakeSlant=0.2,
]

\newcommand{\ncommand}[3]{%
	\expandafter\newcommand\csname #1\endcsname[#2]{#3}%
}
\ncommand{tel}{1}{\begin{telugu}#1\end{telugu}}
\ncommand{dev}{1}{\begin{sanskrit}#1\end{sanskrit}}

\title{
{\Large On the First Computer Science Research Paper in an Indian Language} \\
{\large and the Future of Science in Indian Languages}
}
\author{
Siddhartha Visveswara Jayanti\\
\tel{సిద్ధార్థ విశ్వేశ్వర జయంతి}\\
Dartmouth College\\
Hanover, NH, USA
}
\date{\today}

\begin{document}
\maketitle
\begin{abstract}
    I describe my experience writing the first original, modern Computer Science research paper expressed entirely in an Indian language. 
    The paper is in Telugu, a language with approximately 100 million speakers.
    The paper is in the field of distributed computing and it introduces a technique for proving epistemic logic based lower bounds for multiprocessor algorithms.
    A key hurdle to writing the paper was developing technical terminology for advanced computer science concepts, including those in algorithms, distributed computing, and discrete mathematics.
    I overcame this challenge by deriving and coining native language scientific terminology through the powerful, productive, Pāninian grammar of Samskrtam.   
    The typesetting of the paper was an additional challenge, since mathematical typesetting in Telugu is underdeveloped.
    I overcame this problem by developing a Telugu XeLaTeX template, which I call \tel{తెలుగు}\TeX (TeluguTeX).
    Leveraging this experience of writing an original computer science research paper in an Indian language, I lay out a vision for how to ameliorate the state of scientific writing at all levels in Indic languages---languages whose native speakers exceed one billion people---through the further development of the Sanskrit technical lexicon and through technological internationalization.
\end{abstract}

\section{Introduction}
\label{sec:introduction}

Language remains a great barrier to scientific learning, scholarship, and dissemination in spite of several commendable large scale efforts to promote scientific and technological education across the world.
This is particularly true in India, the world's largest country by population with over 1.4 billion citizens, constituting over 17\% of the world's population---the vast majority of whom do not speak English \cite{IndiaPopulationByBiAndTriLingualism}.
While India has a rich and living tradition of contributions to computation, mathematics, and science, expression of new scientific discoveries has largely been limited to English in recent centuries.
As an illustrative example, the country's top technical universities, including the Indian Institutes of Technology (IITs) and the Indian Institute of Science (IISc) operate almost exclusively in English both in undergraduate and graduate level teaching and in scholarly scientific publications.
Naturally, this status quo disadvantages the vast majority of Indians, who do not speak English, both in scientific access and aspiration.
On the positive side, enabling scientific access and expression in Indian languages can be a great catalyst to technical education for many and a source of satisfaction even for those who speak English. 

In this article, I relate my experience writing the first modern computer science research paper in an Indian language.
I describe the paper, its technical impact, my motivations for writing it, the hurdles I encountered in the process, and my thoughts on the future of scientific writing in Indian languages.

The paper is in Telugu, titled \begin{telugu}సామాన్య జాగృతి-పరిష్కారం\end{telugu}, pronounced \emph{sāmānya jāgṛti-pariṣkāraṃ}, which translates to {\em Resolving Generalized Wake-Up} \cite{జాగృతిపరిష్కారం}.
It pertains to the field of Distributed Computing and is publicly available on {\em ar$\chi$iv} in Telugu under the English title {\em Generalized Wake-Up: Amortized Shared Memory Lower Bounds for Linearizable Data Structures} \cite{teluguarticle}.
This paper mathematically analyzes concurrent algorithms through the lens of knowledge and information propagation, and contributes novel lower bounds on the time complexity of several multicore data structures, including concurrent stacks, queues, and counters.
I also published these results, in their original Telugu form, along with a Sanskrit abstract, as one of the fourteen chapters of my MIT Ph.D. dissertation \cite{SiddharthaPhDThesis}.
In the dissertation, I also translated the results into English in the preceding chapter for the convenience of my readers and thesis committee---Julian Shun (MIT), Charles Leiserson (MIT), and Robert Tarjan (Princeton).

The dissertation won the {\em 2023 ACM Principles of Distributed Computing Doctoral Dissertation Award}.
The award citation summarizes the impact of this Telugu chapter, remarking that,
``Dr.~Jayanti’s PhD dissertation introduces the Samskrtam Technical Lexicon Project, which incorporates ideas from Panini’s generative grammar to facilitate the coining of new technical vocabulary and increase the availability of scientific education and literature in Indian and other world languages. As part of the project, he uses Sanskrit roots to coin words for several concepts in algorithms and multiprocessors in Telugu, and contributes the first modern computer science research paper in the Telugu language, which has about 100 million speakers around the world.''

The results in this Telugu-language research paper have also informed further research in distributed and parallel computing.
In a joint research effort with Princeton's Robert Tarjan, and my MIT colleague Enric Boix-Adserà, we leveraged the results from the paper to design fast algorithms for a concurrent version of the famous disjoint set union data type that nearly meet the task's intrinsic mathematical speed-limits;
our results were published at the Symposium on the Principle of Distributed Computing (ACM PODC) \cite{JayantiTarjanBoix} and in the journal on Distributed Computing \cite{JTjournal}---two premier international venues for distributed computing research.
MIT researchers independently confirmed that our algorithms are among the fastest, in practice, for solving connected components on huge graphs, such as road networks and the Internet \cite{GpuUnionFind, dhulipala2020connectit}; these algorithms are now used in Google's open-source graph-mining library to power ``clustering algorithms which scale to graphs with tens of billions of edges'' \cite{GoogleGraphMiningGithub}.

My path to writing this research paper in Telugu started as a personal journey. 
I am bilingual in Telugu and English, having been born in New Hampshire and grown up in a Telugu speaking family in America.
While my formal education was fully in English, I have long had a proclivity for poetic expression in Telugu.
This fondness for reading and writing poetry developed into a broader study of Telugu and Sanskrit---a classical language of India that bears a similar relationship to Telugu as Latin does to English.
In my undergraduate days at Princeton, I studied Sanskrit as my foreign language.
My dual interest in mathematics and Indian languages opened the door to interesting learnings, including through some research work with Manjul Bhargava of Princeton's Math Department in studying mathematical aspects of Sanskrit grammar and prosody through ancient works on lingusitics and mathematical patterns in poetry.
Equipped with this study, I was initially propelled to write some of my scientific thoughts and discoveries in Telugu as a form of poetic self-expression.
Whereas poetic expression is often used to capture abstract feeling and evoke emotion, scientific writing in Telugu required a concise and precise style to capture technical definitions, algorithms, theorems, and proofs.
It was an exciting challenge.

Simultaneously, my passion for classroom teaching and education led me to see a larger need for making science available in Indian languages.
I have taught computer science and mathematics to students of various levels in America, including in schools, universities, and recreational mathematics and robotics clubs. 
I also take opportunities in my several trips to India to teach over there.
Along with visiting top-ranked Indian institutions, such as the IITs, ICTS, and Microsoft Research India, I have also had the opportunity to visit smaller colleges and schools in the states of Telangana, Andhra Pradesh, and Karnataka---ranging from small village schools to elite private schools in major cities. 
In many of these places, I have directly witnessed the hurdle that English can be to learning science, especially for students who know it only in the capacity of a second or third language.
While I am a first-language English speaker, I can relate to their struggle by imagining what life would be like if I had grown up much the same in New Hampshire, but going to a school where science was taught to me in French (my second language in school). 
Would I have become a computer scientist in such a world? Unlikely.

My personal and anecdotal observations about the need for Indian-language efforts in science match quite readily with the observations of India's Ministry of Human Resource Development. 
India recently undertook a gargantuan initiative to write a new National Educational Policy (NEP 2020), addressing needs for the entire country from pre-K to Ph.D. and beyond. 
The research done in this initiative reveals that “textbooks (especially science textbooks) written in India’s vernaculars at the current time are generally not nearly of the same quality as those written in English” \cite{NEP} and that “students [are] going to school to classes that are being conducted in a language that they do not understand, causing them to fall behind before they even start learning” \cite{draftNEP}. 
Addressing this issue is a massive undertaking, but ameliorating this situation will enable people of diverse backgrounds to pursue STEM and facilitate their inclusion in the scientific research community and in the development of technology.
In this context, my initiative to pen a computer science research paper in Telugu serves as a demonstration of how even the highest echelon of science---the discovery and exposition of new ideas---can be conducted in Indian languages.

A principle barrier to expressing computational and mathematical ideas in Telugu was the lack of technical vocabulary to capture various technologies and scientific concepts that have been studied since the birth of modern computer science.
For example, there was no established vocabulary to talk about ``shared-memory multiprocessors'', ``algorithmic time complexity'', or ``upper and lower bounds''.
So, to write this paper, I had to coin Telugu terms for several technical words that arise in the fields of algorithms, discrete mathematics, and distributed computing.
The productive grammar of Sanskrit \cite{Ashtadhyayi} served as an incredibly powerful tool in deriving such technical vocabulary.
In fact, due to the prevalence of Sanskrit vocabulary across Indian languages, the vocabulary that I derived can easily be used across Indian languages. 
I illustrate this power of Sanskrit grammar in coining concise and precise scientific terminology that can be used seamlessly across Indian languages later in this article.

The lack of a centralized technical lexicon for scientific terminology is one of the big factors that is holding back scientific literature across Indian languages.
I discuss how the approach that I used can be extended across the sciences and across the spectrum of Indian languages to amass a shared scientific lexicon for Indian languages. 
In particular, I propose the building of a Sanskrit based technical lexicon to aid and accelerate original scientific text writing and scientific text translation into Indian languages.
I name this effort the {\em Samskrtam Technical Lexicon Project}.

A secondary hurdle to writing the paper came at the stage of typesetting, particularly because the Telugu language and its associated script are not yet well-supported in typing and type-setting.
Thankfully, I was able to overcome this problem by finding and customizing a suite of disparate tools, including the Lipika input method editor (IME) for typing Telugu and devanagari characters \cite{lipika}, the Kohinoor Telugu font for rendering rare Telugu letters, and XeLaTeX \cite{XeLaTeX} for typesetting.
The tools required extensive configuration and some hacking to get the full set of Telugu characters to type and render correctly, particularly in special environments, such as in mathematical equations.
In my dissertation, I termed the final template files I created for typesetting Telugu scientific text \begin{telugu}తెలుగు\end{telugu}\TeX---i.e., {\em TeluguTeX} \cite{SiddharthaPhDThesis}.
The difficulties I encountered in the process of typesetting suggest the need for further efforts towards internationalization of technology to support languages with non-latin scripts.
I am hopeful that experts of typesetting will take over the development of TeluguTeX and develop similar efforts for other languages and scripts. 

The remainder of this paper consists of five sections. 
I start with a background (Section~\ref{sec:background}), where I give a very brief overview of Indian languages, the associated history of mathematics and computing, and the need for Indian language science texts starting at the school level.
I subsequently discuss the technical content and impact of the Telugu-language computer science paper that I wrote (Section~\ref{sec:technical-result}).
With this background, I explain how I overcame the hurdle of limited pre-coined technical vocabulary in Telugu in algorithms and distributed computing by deriving appropriate technical vocabulary through Sanskrit (Section~\ref{sec:vocabulary}).
I then expand the scope of the discussion to the future of scientific expression in Indian languages, and discuss how my approach can be replicated through the expansion of the Sanskrit technical lexicon (Section~\ref{sec:stlp}).
I also touch briefly on the need for greater efforts towards technological internationalization and the potential role that AI can play in machine translation once human experts blaze the trail.
I end the article with some final remarks (Section~\ref{sec:conclusion}).

\section{Background}
\label{sec:background}

While the {\em Sāmānya jāgṛti-pariṣkāraṃ} work is, to my knowledge, the first {\em modern} computer science work originally composed in an Indian language, it is by no means the first work on computing in an Indian language.
In fact, India has an ancient and deep history of contributions to computing and mathematics.
In this section, I give a brief introduction to Indian languages, mention a few foundational works on computing that exemplify the deep contributions that India has historically made to science through its native tongues, and touch on the need for Indian language literature on science in the modern era and some efforts that have made progress in this direction.

\subsection{The languages of India}

India is the world's largest country by population with an estimated 1.5 billion people.
Indians speak hundreds of native languages, including the country's 22 official languages, which are enumerated in the 8th schedule of the Indian constitution.
Ten of these languages have over 50 million speakers \cite{Ethnologue}.
Several of these languages have large numbers of speakers outside of India, including in neighboring South Asian countries, and throughout the world.

Telugu is one of India's official languages of India enumerated in the 8th schedule. 
It has over 80 million first language speakers in India, and around a 100 million total speakers \cite{2011-India-Census}.
Most of these speakers live in the states of Andhra Pradesh and Telangana, where Telugu is designated an official language of the state.
Additionally, Telugu is also the fastest growing language community in the United States, especially due to the large numbers of Telugu speakers that immigrate to the US to work in the tech industry \cite{BBCTeluguGrowth}.

Sanskrit, (endonym: {\em Saṃskṛtam}), is a classical language of India.
Several Indian languages, including Hindi---the country's largest mother tongue and the world's fourth most spoken first language---are linguistic descendants of Sanskrit. 
Almost all Indian languages, including those which are not direct descendants of Sanskrit, extensively borrow vocabulary, especially technical vocabulary, from the classical language.
Thus, as a progenitor of vocabulary, Sanskrit bears a relationship with Indian languages that is similar to that of Latin and Ancient Greek with English and other European languages.
Sanskrit is also one among the 22 official languages of India enumerated in the 8th schedule, and an official state language in the Indian states of Himachal Pradesh and Uttarakhand.

English also holds an official status in India, as one of the languages of the Indian government, yet only around 10\% of Indians speak it with proficiency as a first, second, or even third language \cite{IndiaPopulationByBiAndTriLingualism}.
Nevertheless, English is the language in which most technical education is conducted at the country's top science and technological universities, including the Indian Institutes of Technology (IITs) and the Indian Institute of Science (IISc).

Indian languages are generally written in their corresponding Indian scripts.
Telugu is written in Telugu Script; the word Telugu is written \tel{తెలుగు} in its native script. 
Today, Sanskrit is most commonly written in a script called Devanagari, although it continues to be and was historically written in many scripts, including for example Telugu Script.
Devanagari is the most widely used Indian script.
It is the primary script of several other languages such as Hindi and Marathi.
The word Sanskrit in its native language and script is written \dev{संस्कृतम्}.
Since this article is in English, I express Telugu and Sanskrit words in Latin script---the script used for English---as prescribed by the ISO 15919 and International Alphabet of Sanskrit Transliteration (IAST) schemes.
The schemes use accented Latin characters to help disambiguate Indian letters that map to the same Latin character.
For example, the words for Telugu and Sanskrit are transliterated to {\em telugu} and {\em saṁskr̥tam}, respectively.


\subsection{Scientific Expression in Indian languages}

While English has been adopted as the primary language of scientific expression in recent centuries, in colonial and post-colonial India, Indian languages were used for this purpose historically.
India has a rich history of contributions to science and computation, and these contributions have traditionally been expressed in Indian languages, generally with a heavy influence of Sanskrit.
Many books have been written on the history of science and mathematics in India (e.g. \cite{kamble2022imperishable}).
Here, I simply list some famous and celebrated examples of Indian contributions that have been foundational in the evolution of computation:
\begin{itemize}
    \item
    {\em Zero and the Hindu Numerals:}
    India is widely recognized as the birthplace of two of the great pillars of modern mathematics and computation: (1) the notion of zero as a number and (2) the decimal number system \cite{devlin2017much}.
        
    \item 
    {\em Computational Grammar}: 
    In a work called the {\em Aṣṭādhyāyī} (circa 500 BCE) \cite{Ashtadhyayi}, Pāṇini described the formal grammar of the Sanskrit language in 3959 rules, called {\em sūtras}.
    In modern terms, the rules amount to the steps of a non-deterministic algorithm which outputs well-formed Sanskrit words given inputs such as a verbal root and a tense, person, and number to conjugate it in.
    The rules themselves are written in a combination of Sanskrit and a Sanskrit-based formal meta-language that Pāṇini invents in the text.
    The use of the formal language, which is a precursor to modern programming languages, yields a succinct and rigorous description of the grammar.

    \item
    {\em Combinatorics through Prosody}:
    The science of prosody---known as {\em Chandaḥśāstra} in Sanskrit---is one of the Vedāṅgas, i.e., traditional disciplines of study in ancient India.
    Several works---such as Piṅgala's {\em Chandaḥśāstra} \cite{Chadassastra} (circa 3rd-2nd century BCE) and Virahānka's {\em Vṛttajātisamuccayā} \cite{vrttajatisamucchaya} (circa 700 CE)---in this discipline analyze poetic meters and characterize utterances by their rhythm.
    In particular, a poetic utterance's rhythm is a binary string, where the two symbols represent the lengths of the syllables in the utterance, either: {\em laghu} (short: one beat of time) or {\em guru} (long: two beats of time).
    In their analyses, Piṅgala and Virahānka derive several mathematically interesting algorithms, including for:
    \begin{itemize}
        \item 
        computing the Fibonacci numbers---since the $n$th fibonacci number is the number of rhythmic utterances that take a total of $n$ beats of time to utter.
        \item
        computing the values of the {\em Piṅgala Méruprastāra} (i.e., Pascal's triangle)---since the value of $n$ choose $k$ is the number of binary strings of length $n$ that contain $k$ laghus.
        \item
        converting from binary to decimal and back---to give a unique numbering of binary strings, which correspond to rhythmic utterances.
    \end{itemize}
    These works are written in Sanskrit and in Prakrit, i.e., a vernacular dialect derived from Sanskrit.
        
    \item
    {\em Infinite Series}: 
    In his text, {\em Yuktibhāṣā} \cite{jyesthadeva2008yuktibhasa}, Jyeṣṭhadeva (1500 - 1575 CE) relates mathematical discoveries of the Kerala School of Astronomy and Mathematics, founded by Mādhava of Sangamagrāma (1340 – 1425 CE).
    Their work originates the study of infinite series.
    For example, one important result of the work is the first infinite series to be discovered which can be used to compute the value of $\pi$.
    This series is now commonly known as the Mādhava-Leibniz series:
    $$\pi/4 = \sum_{n = 1}^\infty \frac{(-1)^{n-1}}{2n - 1}$$
    The {\em Yuktibhāṣā} is originally written in Malayalam with technical vocabulary derived from Sanskrit. 
    Malayalam is a living language with a large population of speakers, mainly centered in southern India, but also with significant populations in West Asian countries.
\end{itemize}

There have been several efforts to revive scientific expression in Indian languages, especially in the last century.
These efforts have largely focused on coining native vocabulary for scientific and technological concepts and translating existing scientific works into Indian languages.
Some examples of the former, include: 
the work of Raghu Vira who coined Indian vocabulary for various molecules and compounds using Sanskrit as part of his {\em Great English-Indian Dictionary} \cite{vira1944}; 
Shrinivasa Varakhedi's {\em English-Sanskrit Computer Dictionary} \cite{varakhedi2020}, in which he coins Sanskrit words for terms relating to computers and their use;
and the work of the {\em Bharatiya Bhasha Samiti}, which produces glossaries of Indian language alternatives for scientific terminology in English.
These works aim to aid in translating modern scientific texts from English into Indian languages.

\subsection{Need for Indian Language science texts at the school level}

It is noteworthy that scientific expression in Indian languages---at all levels of the scholarly process---is in need of great improvement.
The Indian government recently undertook a gargantuan initiative to overhaul the educational policy at all levels---from preschools to universities---to better serve a billion people.
One of their major findings was that education, in general, and science in particular, are easier for people to grasp in their mother tongues, yet, several hundreds of millions of people in India alone are currently unable to receive a quality education in their mother tongues due to lack of educational resources in their native languages \cite{NEP}.
The new educational policy observes that there are ``students going to school to classes that are being conducted in a language [English] that they do not understand, causing them to fall behind before they even start learning,'' and that ``textbooks (especially science textbooks) written in India’s vernaculars at the current time are generally not nearly of the same quality as those written in English. 
It is important that local languages, including tribal languages, are respected and that excellent textbooks are developed in local languages, when possible, and outstanding teachers are deployed to teach in these languages'' \cite{draftNEP} (P4.5.0).
In short, the unavailability and under-availability of scientific texts and scholarship in Indian languages is disadvantaging hundreds of millions of children.
On the positive side, this means that making science available in Indian languages can have a positive impact on a large part of the world's population.

\section{The Technical Impact of the \emph{Sāmānya jāgṛti-pariṣkāraṃ} Paper}
\label{sec:technical-result}

While cutting-edge contributions to mathematics and computing in Indian languages were commonplace until just a few centuries ago, the colonial era changed the scene so dramatically that while India continues to be a top contributor to computational science, expressing these contributions in Indian languages has become unimaginable to most.
Unlike most countries around the world, computer science is {\em de facto} assumed to be a subject that cannot be accessed in native languages. 
This reality is also reflected in the top science and engineering research universities in the country, which operate exclusively in English.
Thus The \emph{Sāmānya jāgṛti-pariṣkāraṃ} paper in Telugu breaks thorough this invisible barrier by making the first modern research contribution to computer science in an Indian language.
Furthermore, its contribution has been prominent in distributed computing, that too not only its academic field, but also touching industry applications.
I explain the contributions of this paper and their impact in this section.

The fundamental object of study in the \emph{Sāmānya jāgṛti-pariṣkāraṃ} paper is the efficiency of algorithms for modern multiprocessors.
The paper thus builds on recent results in distributed computing and algorithms. 
The work establishes a fundamental lower bound on the rate of knowledge propagation in a shared-memory systems, and through reductions, leverages this result to show lower bounds on the time complexities of several concurrent data structures, including: queues, stacks, priority queues, and disjoint set union objects. 

The paper works in the standard asynchronous model of shared-memory computation, where the speeds of the various processes can be vastly different and variable.
Such a system is modeled as if an adversarial scheduler decides when each process executes the next step of its algorithm.
The analysis in the paper centers around the so-called {\em generalized wake-up problem}, which asks the question: how efficiently can a given process gain the knowledge that at least $k$ other processes in the asynchronous system have ``woken up'', i.e.,  executed at least one step of their algorithm?

The queue, stack, and priority queue lower bounds have a logarithmic flavor, and show that while these objects require just constant time per operation when designed for a single process, they require at least $\Omega(\log p)$ time per operation when designed for $p$ processes.
Due to the strength of the underlying knowledge propagation structure, the lower bounds are very strong and cannot be breached by amortization or randomization.
In fact, I show that the results are even robust to approximation, by showing that lower bounds even hold for relaxed definitions of queues and stacks.

The results have had a particularly profound impact on the development of efficient concurrent disjoint set union data structures.
A disjoint set union data structure essentially solves the {\em connected components} problem on graphs under vertex and edge additions.
In other words, it allows a user to incrementally specify the entities and connections in a network and dynamically query whether two nodes in the network are connected.
For example, such a data structure can help answer questions like: can you travel from $x$ to $y$ by train in a rail network? or are two people, Alice and Bob, connected through mutual friends in a social network?
Due to the prevalence of graph structured data, and the humongous size of modern social network, web, and road graphs, a recent study has found that computing connected components is actually the largest industrial application of graph algorithms \cite{DBLP:journals/vldb/SahuMSLO20}.

In a joint research effort with Robert Tarjan and Enric Boix-Adserà, we leveraged the results of the \emph{Sāmānya jāgṛti-pariṣkāraṃ} paper and designed concurrent disjoint set union algorithms that near the task's intrinsic mathematical speed-limits, with an $O\left(\alpha{\left(n, \frac{m}{np}\right)} + \log{\left(\frac{np}{m} + 1\right)}\right)$ time complexity per operation, when a total of $m$ operations are performed by $p$ processes on a data structure with $n$ nodes.
Here $\alpha(\cdot, \cdot)$ is the extremely slow growing inverse-Ackermann function, which increases in its first argument and decreases in its second argument. 
The function grows so slowly, that $\alpha(n, 1)$ is just four, when $n$ is the estimated number of atoms in the observable universe \cite{clrs}.
Thus, our algorithms' work complexity scales only {\em logarithmically} with the number of processors and is thus the first concurrent set union algorithm to achieve {\em almost linear speed-up} in the number of processors.
Our results were published at {\em the Symposium on the Principle of Distributed Computing} and in the journal on {\em Distributed Computing}---two premier international venues for distributed computing research \cite{JayantiTarjanBoix, JTjournal}.
MIT researchers independently confirmed that our algorithms are the fastest in practice for solving connected components on huge graphs, such as the Internet and road networks \cite{GpuUnionFind, dhulipala2020connectit}.
These algorithms are now used in Google's graph-mining library for ``clustering algorithms which scale to graphs with tens of billions of edges'' \cite{GoogleGraphMiningGithub}.

The results from the Telugu research paper also constitute a part of my Ph.D. dissertation, where the results appear in their original Telugu with an additional abstract in Sanskrit, and in English translation.
The dissertation won the {\em ACM Principles of Distributed Computing Doctoral Disseration Award} in 2023.
The award citation summarizes the technical contributions of these chapters by noting that ``Dr. Jayanti defines a generalization of the fundamental wake-up problem, permitting him to prove fundamental new hardness results for many standard data structures, including queues, stacks, priority queues, counters, and union-find data structures.''

\section{Deriving the Vocabulary}
\label{sec:vocabulary}

Now that I have explained the results, imagine how they would be written in a research paper, but with one catch: you must write them in a langauge that has no previous vocabulary for several of the technical terms, including: {\em time complexity}. {\em lower bounds}, {\em asynchronous}, and {\em shared-memory multiprocessor}.
That is the difficulty faced by a computer scientist attempting to express research in an Indian language: a dearth of the rich technical vocabulary needed to express recent scientific advances.

I overcame this need for technical terms using an age-old method used across Indian languages---deriving vocabulary through the generative grammar of Sanskrit.
Paninian grammar has been around for over two millenia and has been the method of preference for deriving technical vocabulary in Indian languages over the centuries.

I first explain the basic principles behind such vocabulary generation by giving simple examples.
Then, I explain how long and sophisticated compound words can be derived, through the example of the word: {\em shared-memory multiprocessor} in the next subsection.

\paragraph{Example term derivations}
Terms for many specialized technical words that are not readily available in dictionaries can be derived very easily.  
I list just a few examples of such words from concurrent, parallel, and distributed computing that I derived for the paper below.

\begin{itemize}
    \item 
    \underline{{\em asamakālika} (\begin{sanskrit}असमकालिक\end{sanskrit}) for `asynchronous'}:
    the English word asynchronous means `not at the same time', and it is used technically to describe certain schedules of multiple processors executing in a larger computational system. 
    The Sanskrit term {\em asamakālika} is derived from {\em a-} (not), {\em sama} (same), {\em kāla} (time).
    Incidentally, the English term `asynchronous' also comes from combining the Greek morphemes {\em a-} (not), {\em syn-} (together), {\em khronos} (time).

    \item
    \underline{{\em kālamātrā} (\begin{sanskrit}कालमात्रा\end{sanskrit}) for `discrete time step'}:
    in computer science, a distributed system is often modeled as running in discrete time steps.
    The Sanskrit term {\em kālamātrā} is derived by combining {\em kāla} (time) with {\em mātrā} (a measurable unit).
    In fact, the term {\em mātrā} has been used for millenia in Indian prosody when measuring lengths of syllables in poetry \cite{Chadassastra, vrttajatisamucchaya}.

    \item
    \underline{{\em ripukālabandha} (\begin{sanskrit}रिपुकालबन्ध\end{sanskrit}) for `adversarial schedule'}:
    in distributed computing, an {\em adversarial schedule} refers to a particular worst possible execution of various threads as they execute an algorithm together.
    The Sanskrit term {\em ripukālabandha} is derived by combining {\em ripu} (adversary), with {\em kālabandha} (schedule).
    The term {\em kālabandha} itself comes from {\em kāla} (time) and {\em bandha} (bind), since a schedule binds one's time.    
\end{itemize}

Respecting the grammatical rules of Telugu \cite{AndhraSabdaChintamani}, which borrow Sanskrit words and often add a {\em -mu} suffix to form the nominative case of a masculine or neuter word or shorten the trailing vowel for feminine words, these terms are adapted as {\em asamakālikamu} (\begin{telugu}అసమకాలికము\end{telugu}), {\em kālamātra} (\begin{telugu}కాలమాత్ర\end{telugu}), and {\em ripukālabandhamu} (\begin{telugu}రిపుకాలబంధము\end{telugu}) in the Telugu paper \cite{జాగృతిపరిష్కారం}.

\paragraph{Deriving a complex technical term}
Even complex technical words that refer to modern technology can be derived using Pāṇini's Aṣṭādhyāyī generative grammar of Sanskrit. 
As described above, these words can be readily adapted into other Indian languages using grammar techniques that have been natively developed by these languages over millenia.

I will now demonstrate how Sanskrit's derivational grammar can be used to coin technical vocabulary for even very specialized and complex words in the sciences, and how these words can then be adapted to a different Indian language to be used in practice.
In the paper \cite{జాగృతిపరిష్కారం}, I had to talk about a ``shared-memory multiprocessor'', i.e., a computing system that has a CPU with multiple processors, which all share a common random access memory.
I derived the term ``{\em saṁvibhaktasmr̥ti bahusaṁsādhakamu}'' (Telugu: \begin{telugu}సంవిభక్తస్మృతి బహుసంసాధకము\end{telugu}) for shared-memory multiprocessor.
Here is the derivation:

\vspace{0.1in}
\emph{saṁvibhakta-smr̥ti bahusaṁsādhakamu =}

\emph{\hspace{0.1in} ((saṁ $+$ vi $+$ bhaj $+$ kta) $+$ (smr̥ $+$ ktin)) $+$ ((baṁh $+$ ku) $+$ (saṁ $+$ sādh $+$ ṇvul)) $+$ mu}
\vspace{0.1in}

The derivation may look intimidating at first, but looking at it in small chunks makes it easier to understand and absorb. 
Let me explain it piece by piece.

The Sanskrit root \emph{bhaj} means `to give'. 
Modifying it with the prefix \emph{vi-} forms \emph{vibhaj}, meaning `to partition'.
Further modifying it with the prefix \emph{saṁ-} forms \emph{saṁvibhaj}, meaning `to share'.  
By adding the suffix \emph{-kta} to it, we get \emph{saṁvibhakta}, meaning `shared'.

\emph{smr̥} is the Sanskrit root meaning `to remember, recollect, or memorize'. 
By adding the suffix \emph{-ktin} to it, we get \emph{smr̥ti}, meaning `memory'.
Compounding \emph{saṁvibhakta} (shared) with \emph{smr̥ti} (memory) gives us \emph{saṁvibhaktasmr̥ti} (shared-memory).

Combining the root \emph{baṁh} `to grow' with the suffix \emph{-ku}, gives us \emph{bahu} (many or multi). 
Modifying the root \emph{sādh} `to accomplish' with the prefix \emph{saṁ-} forms \emph{saṁsādh} (to process); adding the suffix \emph{-ṇvul} gives us \emph{saṁsādhaka} (processor). 
Compounding \emph{bahu} (multi) with \emph{saṁsādhaka} (processor) yields \emph{bahusaṁsādhaka} (multiprocessor).

Compounding \emph{saṁvibhaktasmr̥ti} (shared-memory) with \emph{bahusaṁsādhaka} (multiprocessor) gives us \emph{saṁvibhakta-smr̥ti bahu-saṁsādhaka} (shared-memory multi-processor) in Sanskrit. 
Adapting this term derived in Sanskrit to another Indian language is simple.
In the paper, the ending \emph{-mu} is added to adapt the word to Telugu, giving us \emph{saṁvibhakta-smr̥ti bahu-saṁsādhakamu} (shared-memory multi-processor).

The derivations above illustrate the power of the Sanskrit grammar to seamlessly derive modern technical terms.
They also demonstrate the ease with which these terms can be adapted to other Indian languages, such as Telugu.

\section{On the Future of Science Expression in Indian Languages}
\label{sec:stlp}

The future of scientific expression in Indian language will likely rely on many factors, including technical linguistic factors, governmental and institutional support, technological support, and economic factors.
It is hard to imagine large-scale scientific expression in Indian languages becoming possible without educational institutions supporting high-quality teaching of science in these languages, and this would likely require a serious effort from the government and scientific and academic establishment.
Similarly, students who study science in their mother-tongues would need to find means to succeed in the job market.
This would require companies that are willing to hire those who study science in their native language or modalities for prospective employees to transfer their scientific learning to an English-medium work environment.
These significant institutional and economic factors are likely to be a focus of the Indian government, given the initiative it has shown through its reforms in the National Education Policy.
I focus here on the technical side of the challenge.

The {\em technical lexicon}---consisting of precise scientific terms that refer to well defined technical concepts---is a keystone to scientific education, communication, and innovation.
Thus, developing a technical lexicon for Indian languages is imperative in order to improve science education, science-availability, and ultimately enable the expression of new scientific discoveries for a large part of the Indian---and thereby world's---population.

\subsection{The Samskrtam Technical Lexicon Project}

The {\em Sāmānya jāgṛti-pariṣkāraṃ} paper demonstrates that even sophisticated technical vocabulary can be derived through Sanskrit's generative grammar.
I believe that furthering this effort to expand the scientific expressivity of Indian languages through Sanskrit derivations, which I termed the {\em Samskrtam Technical Lexicon Project (STLP)} in my Ph.D. thesis \cite{SiddharthaPhDThesis}, is a great technique for establishing an Indian language scientific lexicon, not only for computer science, but also across the sciences.
Coining technical terms using the productive grammar of Sanskrit has several advantages, including: significantly increasing science-availability across India, enabling precise scientific expression, providing a unified technical vocabulary across Indian languages, and respecting the incredible linguistic diversity and vernaculars across India.
I detail some of these advantages below.

\begin{enumerate}
    \item
    {\em Expanding science-availability.}
    New scientific and technical vocabulary is currently a principal barrier to enabling the writing of scientific texts in many languages, ranging from textbooks in grade schools to scholarly research papers.
    Developing a shared Sanskrit vocabulary for technical terms would ameliorate this barrier for languages across the country simultaneously, and expand science-availability in language communities that span a billion people.

    \item
    {\em Precise scientific expression.}
    The productive grammar of Sanskrit makes it possible to methodically derive more complex technical terms from well-known simple parts, thereby producing vocabulary that is very well suited for clear, concise, and precise scientific expression.
    Sanskrit has an ancient yet still thriving tradition of formal grammar, expressed in rules of Pānini's celebrated Ashtādhyāyī \cite{Ashtadhyayi}.
    The algorithmic structure of this grammar ensures that the morphological form of a word matches its semantic meaning.
    In other words, larger more complex words are methodically constructed from meaningful smaller parts according to precise rules.
    Thus, Sanskrit technical words precisely express the concept they are meant to capture as I described in the previous section.
    Additionally, since more complex terms can be broken down into smaller parts that occur in familiar words, these terms are easier to learn for a new student.

    \item 
    {\em Unifying vocabulary that respects linguistic diversity.}
    Sanskrit is a prominent source of technical vocabulary not only for Indian languages, but also for innumerable languages across South Asia and the rest of the world.
    Thus, terms derived in Sanskrit serve as a unifying vocabulary across India, its neighbors, and the greater world.
    The various regions of India have natively adopted, over centuries and millenia, ways to adapt words from the Sanskrit lexicon into their native languages. 
    As a result, Sanskrit terms can be used by the different languages across India in a way that respects their linguistic identity and the country's linguistic diversity, by adapting the form of the word accordingly.
    For example, the Sanskrit word {\em saṅgaṇaka}, meaning computer, may be pronounced {\em saṅgaṇak} in Hindi, {\em saṅgaṇakamu} in Telugu, and {\em saṅgaṇagam} in Tamil.
    Nevertheless, in the same way that an English speaking scientist can collaborate easily with a French scientist due to etymological similarities between their languages---for instance, the French {\em intégrale} and the English {\em integral}---speakers of Indian languages can communicate across language barriers with shared Sanskrit vocabulary.    
    \item
    {\em Existence of some important vocabulary.}
    The Samskrtam lexicon already has a significant corpus of scientific terminology even in relatively new sciences like Computer Science (e.g., {\em vidhikalpa} for algorithm).
    Such technical vocabulary can be readily used in any language that inherits vocabulary from Sanskrit---Assamese, Urdu, Odia, Kannada, Kashmiri, Gujarati, Tamil, Telugu, Punjabi, Bengali, Marathi, Malayalam, Nepali, Sinhala, Hindi, and many more.
    (The preceding languages are listed in Sanskrit-dictionary order.)
    Generally, adapting the Sanskrit word to one of these languages requires a minor modification in the ending of the word to match the grammatical endings of the target language. 
    For example, {\em vidhikalpa} becomes {\em vidhikalpamu} in Telugu and {\em vidhikalp} in Hindi.    
\end{enumerate}

\subsection{Internationalization}

In this digital age, technological support for languages has become a corner stone issue for languages to thrive.
Several Indian languages and scripts are not yet well-supported in modern technology.
In fact, I had to spend a considerable effort at the stage of typesetting, since the Telugu language and its associated script pose difficulties to modern keyboards and typesetting systems.
For example: 
the Telugu keyboard on my MacBook Pro laptop was hard to use, since it was missing some of the infrequently used letters in the Telugu alphabet; 
several Telugu fonts that I tried using did not render certain glyphs in compound letters (known as {\em guṇintālu} and {\em samyuktāksharālu} in Telugu) correctly;
and even standard XeLaTeX---an interanationalized version of the commonly used scientific typesetting system LaTeX---did not render general Telugu inputs out-of-the-box.
After much struggle and trial and error, I finally settled on a combination of tools that worked for my purpose of rendering the {\em Sāmānya jāgṛti-pariṣkāraṃ} paper in Telugu.

For my Input Method Engine (IME)---i.e., virtual keyboard that transliterates inputs on my qwerty-keyboard to Telugu characters---I used Lipika \cite{lipika}.
Lipika is a tool that supports a wide array of Indian scripts, making it a useful choice for general Indian language typsetting.
I had to use both Telugu script and Devanagari script (the script commonly used for Sanskrit) in my project, and Lipika had support for both of these.
Another useful feature of Lipika was its customizability.
While Lipika did not support some of the rarer Telugu characters such as \tel{ౘ} and \tel{ౙ} out-of-the-box, I was able to easily customize its mappings to access the full Telugu character set (supported by Unicode).

Even as Lipika allowed me to type all the Telugu characters supported by Unicode, most Telugu fonts do not themselves support all the characters.
Furthermore, the Telugu script is an abugida, meaning that consonant-vowel clusters are rendered as a single combined characters.
Some fonts have only partial support for letters of the Telugu script---meaning that they can render a letter in isolation, but cannot properly combine it with its cluster to produce the resultant character.
After some considerable effort, I came upon a font named Kohinoor Telugu which is the only Telugu font I know of to-date that properly renders all the letters and clusters that I have needed to type.

Even after I settled on an input method and font that worked, it took a considerable amount of customization to render the Telugu in a scientific typesetting system.
In the end, I was able to configure XeLaTeX to render the Telugu characters both in running text and in mathematical formulas, macros, special environments (e.g., theorem statements and algorithmic blocks).
In my dissertation, I termed the final template files I created for typesetting: \begin{telugu}తెలుగు\end{telugu}\TeX---romanized as {\em TeluguTeX} \cite{SiddharthaPhDThesis}.

The difficulties that I encountered in the process of typesetting the paper in Telugu are common to several Indian languages, including others with a large number of speakers.
This suggests the need for further efforts in internationalization of software technology.
I am hopeful that experts of typesetting will take over the development of TeluguTeX and similar efforts for other scripts. 
More generally, Telugu and various other languages continue to suffer from technological under-availability. 
To address this concern, I support making more rapid progress on existing technological initiatives that empower linguistic diversity, such as Unicode \cite{unicode}, XeLaTeX \cite{XeLaTeX}, and Internet Internationalization \cite{internetinternationalization}.

\subsection{A Brief Note on Artificial Intelligence and Machine Translation}

Given that most modern scientific texts are written and disseminated in English, another important dimension in increasing scientific access in Indian language communities lies in translating existing and upcoming scientific works from English.
One way to scale up translation efforts is through the use of generative Artificial Intelligence (AI) tools, such as Large Language Models (LLMs).
While AI tools can be powerful, modern AI is still a system that relies largely on sophisticated pattern recognition; it can neither generate new Indian language scientific terminology by itself nor start translating scientific literature without training on careful translations of several scientific works across domains by experts.
Thus, I believe that vocabulary generation through the Samskrtam Technical Lexicon Project (STLP) and translations of some scientific works across domains will need to be done, at least initially, by human experts.
Hopefully, this initial effort can get rapid returns if appropriate AI systems can be trained to produce future vocabulary and machine translations.

\section{Final Remarks}
\label{sec:conclusion}

India has historically been a world leader in computing and mathematics with lasting contributions like the invention of zero, the decimal number system, binary-decimal conversion, formal grammars, and convergent infinite series.
These discoveries, which span millennia, were all expressed in the native languages of India---mostly in Sanskrit or with the use of Sanskrit vocabulary.
In colonial and post-colonial India, scientific education largely moved to English, but this has been to the detriment of the innumerable non-English speakers of the nation.
In recent times however, there has been an increasing interest in ameliorating this situation, and making science and scientific education available in Indian languages once again.
The importance of mother-tongue education in science is advocated for strongly in India's recent {\em National Education Policy 2020}.
In short, enabling scientific access and expression in Indian languages can be a great catalyst to technical education for many and a source of satisfaction for the entire population of the most populous country of the world.

In this light, one of the main impacts of the {\em Sāmānya jāgṛti-pariṣkāraṃ} paper---the first modern computer science research paper in an Indian language---is in demonstrating the viability of doing cutting-edge research in an Indian language.
The paper is written entirely in Telugu, an Indian language with about a 100 million speakers, and derives new scientific results regarding algorithms for modern multiprocessors.

The {\em Samskrtam Technical Lexicon Project}---which ensues from the Telugu paper---outlines a framework to develop science in all Indian languages that can demonstrably enable scientific expression all the way up to the research frontier.
The project seeks to jump a primary technical hurdle to such expression---i.e., the dearth of technical vocabulary to express recent scientific ideas due to the hiatus in Indian language science research in the past few centuries.
This goal is achieved by using the powerful generative grammar of Sanskrit to derive the technical vocabulary needed to express cutting-edge results in the sciences, including computer science which is among the newest branches of science.
The derived terms are concise and precise, making them suitable for clear scientific expression, exposition, and communication.
This vocabulary can then be used uniformly across Indian languages, at all levels of education (pre-K to Ph.D. and beyond), while respecting the native grammar and rules of regional languages and vernaculars.
Furthermore, only the experts need to understand how to derive words, but the structure in the words is helpful for even elementary school students, since more complex concepts are built from simpler understandable parts.
Once the vocabulary is derived, it can empower over a billion citizens of India and speakers of Indian languages around the world in learning, communicating, and advancing science, engineering, and technology.
To summarize, a Sanskrit based technical lexicon can serve as a powerful conduit to achieving the vision of the National Education Policy of India---educationally empowering the entire nation into this age of information.

Along with the development of the lexicon, several other factors need to fall into place to make the vision of Indian language scientific expression a reality.
These factors range institutional support from governments and employers and technological support for supporting Indian languages and scripts on computers, phones, and the internet.
Focusing on the technological aspect, a lot of ground remains to be covered in terms of internationalizing technology.
Even with internationalization efforts such as Unicode, it still took considerable effort to typeset a scientific manuscript in Telugu.
The difficulties ranged: finding an input method for typing Telugu Unicode characters, lack of support for rarer characters across many Telugu fonts, and inadequate support for Telugu in typesetting systems.
In the case of the research paper, I was able to piece together and reconfigure a suite of existing tools---Lipika, Kohinoor Telugu, and XeLaTeX---which collectively allowed for the creation of the \tel{తెలుగు}\TeX\ template for typesetting scientific documents in Telugu.
The wider problem of internationalization however, remains unsolved.
For Indian languages to be used for science and technology on a larger scale, we will need a concerted effort to internationalize technology.
Of course, this will be helpful not just for Indian languages, but also for all the languages of the world. 

The first computer science research paper in an Indian language opens up the scope for several future lines of work.
It would be excellent to see more scientific research being published across domains of study and across Indian languages.
Another important direction is in translating existing science into these languages. 
Indeed, as efforts to produce scientific research in Indian languages succeed, it will also become important to translate science from Indian languages to English and other world languages.
All of these lines of work naturally benefit from close collaborations between scientists and language experts, and would especially benefit from those who have expertise in both.
I believe such efforts and collaborations will naturally lead to the expansion of the Samskrtam Technical Lexicon; but it will become important to standardize vocabulary in the process---which may happen organically or require a standardization framework.

Along with work at the research frontier, it is important to develop educational curricula that can effectively teach science, technology, engineering, and mathematics (STEM) to students in their native tongues.
Such an effort would benefit from a significant and effective collaboration between scientists, language experts, and educators at all levels of academia.
Such a collaboration which may be initiated by the Indian government.
Naturally, once such an educational curriculum gets underway, it will become important to ensure that those trained in their native languages can find jobs and also interface with the international scientific community.
In the current status quo, this would involve training these new scientists in English and other world languages.

I look forward to seeing more scientific literature developing across different languages at all levels of study---from kindergarten to the research frontier---and I am hopeful that this will open up the joy of the scientific pursuit to all those who seek it.

\bibliographystyle{plain}
\bibliography{resources/ref}

\end{document}